\documentstyle[12pt]{article}

\begin{document}
\title{Null Singularities in Colliding Waves}
\author{ Ozay Gurtug and Mustafa Halilsoy \\
Department of Physics, Eastern Mediterranean University \\
G.Magusa, North Cyprus, Mersin 10 - Turkey \\
email: ozay.gurtug@emu.edu.tr}
\maketitle
\begin{abstract}
{\small Colliding Einstein - Maxwell - Scalar fields need not necessarily
 doomed to
become in a spacelike singularity. Examples are given in which null singularities
emerge as intermediate stages between a spacelike singularity and a regular
horizon.}
\end{abstract}
\newpage
Coupling scalar fields to a static charged black hole (BH) [1,2] (with
the respective Reissner-Nordstrom (RN) and Newman-Janis-Winicour (NJW) limits)
converts its horizons into null singularities is a known fact. This fact was
 rediscovered recently in a related context involving the perturbation
of a charged BH by ingoing pulses of scalar fields [3,4]. It has been shown that the
inner horizon of a RN BH is unstable against such perturbations and transforms it into
a null singularity. This is an interesting development since it would mean that
a  BH upon being slightly perturbed will not find it so simple to act as a gateway to 'other worlds'
. With this example in mind and the similarity in the dynamics of
collapse of a scalar field and colliding Einstein-Maxwell-Scalar (EMS) fields
motivates us to explore analogous singularities in EMS fields.
Ori has already discussed null singularities in plane symmetric spacetimes [5].
 This must be important
because as Tipler [6] has shown the tidal distortion experienced by an infalling object
is small as it hits such a singularity.\\
In this Letter we consider two concrete cases in the field of colliding plane waves
(CPW). Our first case is the collision of two electromagnetic (em) shock waves, known
as the Bell-Szekeres (BS) [7] solution that yields a quasiregular 'singularity' ( this
we consider equivalent to a horizon). In our second example we consider the horizon
forming CPW found by Chandrasekhar and Xanthopoulos (CX) [8] which is isometric to the
region in between the horizons of the Kerr-Newman BH. We show that by choosing appropriate
scalar fields both the quasiregular 'singularity' of the BS spacetime and the horizon
of the CX metric transform into null singularities.\\
Our first line element is the linearly polarized metric
\begin{equation}
ds^2=\Delta ^{1-A}Z^2 \left(\frac{d\tau^2}{\Delta}-\frac{d\sigma^2}{\delta} - \delta dx^2\right)
-\Delta ^A Z^{-2} dy^2 
\end{equation}
where the notation goes as 
\begin{eqnarray}
\tau + \sigma &=&2u\sqrt{1-v^2} \nonumber \\
\tau - \sigma &=&2v\sqrt{1-u^2} \nonumber  \\
\Delta &=& 1- \tau^2  \nonumber \\
\delta &=& 1 - \sigma ^2 \nonumber \\
2Z&=& a(1+ \tau )^A + b (1- \tau )^A 
\end{eqnarray}
in which $ (a,b) $ and $ A $ are constants and $ (u,v) $ are the null coordinates.
We choose $ 0<A<1 $ to represent the scalar charge while $ (a,b) $ stand for the
em parameters. The scalar field is
\begin{equation}
\phi ( \tau )= \frac{1}{2} \sqrt{1- A^2} \ln{ \frac{1+ \tau }{1- \tau }}
\end{equation}
which implies that for $ A=0 $ there would be a background scalar field already and
the singularity is spacelike. As we increase $ A $ toward unity the scalar field
 diminishes and the singularity of the spacetime transforms to the removable
 quasiregular 'singularity'. The singularity at $ \tau = 1 $, however, becomes null for
 $ 0<A<1 $. Since the null coordinates $ (u,v) $ are to be multiplied by the step
 functions $ \theta (u) $ and $ \theta (v) $ apt for the collision problem the
 $ u $- dependent incoming pulse energy density is
 \begin{eqnarray}
 4 \pi T_{uu}= \Phi ^{(0)}_{22} &=& \frac{ \theta (u)}{4Z^2(1-u^2)^2} \left\{ (1-A^2) \left[ b^2(1-u)^{2A}
 + a^2 (1+u)^{2A} \right] \right. \nonumber \\
 & & \nonumber \\
 & & \left. +2ab(1+A^2)(1-u^2)^A \right \}
 \end{eqnarray}
 where $ 2Z=a(1+u)^A +b(1-u)^A $. For $ A=1 $ (and $ a=b $ ) this metric reduces
 to the well-known BS solution. The scale invariant Weyl scalar $ \Psi^{(0)}_{2}
  $ is found to be

 \begin{eqnarray}
 \sqrt{1-u^2} \sqrt{1-v^2} \Psi^{(0)}_{2} &=& \frac{1-A}{ \Delta }+\frac{A}{4Z^2}
 \left \{ a^2(1+ \tau )^{2A-1} + b^2 (1- \tau )^{2A-1} \right. \nonumber \\
 & & \nonumber \\
 & & \left. + 2ab (1-2A) \Delta^{A-1} \right \}
 \end{eqnarray}
 in which the $ \tau =1 $ singularity is manifest. The remaining scalars $ \Psi^{(0)}_{0} $
 and $ \Psi^{(0)}_{4} $ also have similar structure but these latter two have
 in addition an impulsive component. For $ A=1 $ ( and $ a=b $ ) all conformal
 curvatures vanish for $ u>0, v>0 $ and the singularity $ \tau =1 $ becomes removable.
 Thus the null singularity arises as an intermediate stage in between a spacelike
 quasiregular singularity and a horizon.\\
 As a second example we consider the CX metric
\begin{equation}
ds^2=X \left( \frac{d \tau ^2}{ \Delta}-\frac{d \sigma ^2}{ \delta} \right) -
\Delta \delta \frac{X}{Y} dy^2 - \frac{Y}{X} \left( dx - q_{2} dy \right)^2
\end{equation}
where
\begin{eqnarray}
X&=&\frac{1}{\alpha^2}\left[(1-\alpha p \tau)^2+\alpha^2q^2\sigma^2\right] \nonumber \\
Y&=&1-p^2\tau^2-q^2\sigma^2 \nonumber \\
q_{2}&=&-\frac{q\delta}{p\alpha^2} \frac{1+\alpha^2-2\alpha p \tau}{1-p^2\tau^2-q^2\sigma^2} 
\end{eqnarray}
in which the constant parameters $\alpha,p$ and $q$ must satisfy
\begin{eqnarray}
0<\alpha\leq1 \nonumber \\
p^2+q^2=1
\end{eqnarray}
This metric admits a horizon instead of a spacelike singularity at $ \tau=1 $.
We add now a scalar field $ \phi $ satisfying $ \Box \phi = 0 $, or equivalently
$ \left( \Delta \phi_{ \tau } \right)_{ \tau }= \left( \delta \phi_{ \sigma } \right)_{ \sigma }$
as follows [9]: By shifting the metric function $ X \rightarrow Xe^{-\Gamma} $ we
couple the scalar field consistently with Einstein-Maxwell's fields
 where $ \Gamma $ is determined from the line
integral
\begin{equation}
\Gamma = 2 \int{ \frac{ \phi^{2}_{u}}{U_{u}}du} + 2 \int{ \frac{ \phi^{2}_{v}}{U_{v}}dv}
\end{equation}
in which $ e^{-U}= \sqrt{ \Delta \delta } $. Choosing a simple class of scalar
field such as $ \phi ( \tau )= \frac{k}{2} \ln{ \frac{1+ \tau }{1- \tau}} $, with
 $ k=constant $ results in $ e^{- \Gamma}= \left( \frac{1- \tau ^2 }{ \tau ^2 - \sigma ^2} \right)^{k^2} $.
 With the addition of this scalar field we can see from the energy momentum scalar
 $ T^{ \alpha }_{ \alpha } $ and $ T_{ \mu \nu } T^{ \mu \nu } $, which are divergent
 that $ \tau =1 $ is singular. Furthermore, the fact that as $ \tau \rightarrow 1 $ the metric
 function $ g^{\tau \tau } \rightarrow 0 $ for the case $ k^2<1 $ implies that it is
 a null singularity. For $ k^2 \geq 1 $, however, it retains the spacelike character
 which is standard to CPW. Letting $ q=0 $ and using the transformation
\begin{equation}
t=m \alpha x, \hspace{.5cm} y= \phi, \hspace{.5cm} \tau=\frac{m-r}{\sqrt{m^2-e^2}}, \hspace{.5cm}
\sigma=\cos\theta
\end{equation}
with $ m \alpha = \sqrt{m^2 - e^2} $ we obtain
\begin{eqnarray}
ds^2 &=& \left ( 1- \frac{2m}{r} +\frac{e^{2}}{r^{2}} \right )dt^2 -
  e^{- \Gamma } \left ( 1- \frac{2m}{r} +\frac{e^{2}}{r^{2}} \right )dr^2  \nonumber \\
  & & \nonumber \\
  & & - r^2 \left( e^{ - \Gamma } d \theta ^2 + \sin ^2 \theta d \phi ^2 \right)
\end{eqnarray}
which is a spherically non-symmetric extension of the RN metric with a null singular
horizon. Our method generates infinitly many metrics with null ( or directionally
null, depending on the choice of the scalar field) singular horizons that may
find application in BH's.


\newpage
\begin{thebibliography}{99}
\bibitem{} R.Penney, Phys. Rev. {\bf 182}, 1383 (1969).
\bibitem{} A.I.Janis, E.T.Newman and J.Winicour, Phys. Rev. Lett. {\bf 20},
878 (1968).
\bibitem{} L.M.Burko, Phys. Rev. Lett. {\bf 79}, 4958 (1997).
\bibitem{} L.M.Burko, Phys. Rev. {\bf D 59}, 024011 (1998).
\bibitem{} A.Ori, Phys. Rev. {\bf D 57}, 4745 (1998). 
\bibitem{} F.J.Tipler, Phys. Lett. {\bf 64 A }, 8 (1977).
\bibitem{} P.Bell and P. Szekeres, Gen. Rel. Grav. {\bf 5}, 275 (1974).
\bibitem{} S.Chandrasekhar and B.C.Xanthopoulos, Proc. R. Soc.  London {\bf A 
414 }, 1 (1987).
\bibitem{} O.Gurtug and M.Halilsoy, Report No: gr-qc 0006038.
\end{thebibliography}
\end{document}